\documentclass{aamas2017}

\usepackage{times}
\usepackage{helvet}
\usepackage{courier}
\usepackage{amsmath}

\usepackage{hyperref}
\usepackage{enumitem}

\usepackage[usenames, dvipsnames]{color}
\usepackage[table,x11names]{xcolor}
\definecolor{lightgray}{gray}{0.90}
\usepackage{tikz}
\usepackage{pgfplots}

\usetikzlibrary{pgfplots.dateplot}
\usepackage{pgfplotstable}
\usepackage{filecontents}
\usetikzlibrary{backgrounds}
\pgfplotsset{width=9cm}

\definecolor{wsblue}{RGB}{0, 70, 140}
\hypersetup{
     colorlinks   = true,
     citecolor    = wsblue,
     linkcolor    = wsblue,
     urlcolor     = wsblue
}
\begin{document}


\title{Moving Target Defense for Web Applications using\\Bayesian Stackelberg Games}



%
%
%
%

%

\numberofauthors{1}

\author{
%
\alignauthor
Sailik Sengupta, Satya Gautam Vadlamudi, Subbarao Kambhampati\\
       \affaddr{Yochan Group, School of CIDSE}\\
       \affaddr{Arizona State University}\\
       \email{\tt{\{sailiks, gautam, rao\}@asu.edu}}
\and
\alignauthor
Marthony Taguinod, Adam Doup\'e, Ziming Zhao, Gail-Joon Ahn\\
       \affaddr{SEFCOM Lab, School of CIDSE}\\
       \affaddr{Arizona State University}\\
       \email{\tt{\{mtaguino, doupe, zmzhao, gahn\}@asu.edu}}
}

\maketitle

\begin{abstract}
The present complexity in designing web applications makes software security a difficult goal to achieve. An attacker can \textit{explore} a deployed service on the web and attack at his/her own leisure. Moving Target Defense (MTD) in web applications is an effective mechanism to nullify this advantage of their reconnaissance but the framework demands a good switching strategy when switching between multiple configurations for its web-stack. To address this issue, we propose modeling of a real-world MTD web application as a repeated Bayesian game. We then formulate an optimization problem that generates an effective switching strategy while considering the cost of switching between different web-stack configurations. To incorporate this model into a developed MTD system, we develop an automated system for generating attack sets of Common Vulnerabilities and Exposures (CVEs) for input attacker types with predefined capabilities. Our framework obtains realistic reward values for the players (defenders and attackers) in this game by using security domain expertise on CVEs obtained from the National Vulnerability Database (NVD). We also address the issue of prioritizing vulnerabilities that when fixed, improves the security of the MTD system. Lastly, we demonstrate the robustness of our proposed model by evaluating its performance when there is uncertainty about input attacker information.
\end{abstract}

\section{Introduction}
Present day web applications are widely used by businesses to provide services over the Internet. Oftentimes, sensitive business and user data are managed by these applications. Vulnerabilities in these web applications pose serious threats to the confidentiality and integrity of both businesses and users~\cite{nyt}.

There exist numerous static (white-box) and dynamic (black-box) analysis tools for identifying vulnerabilities in a system \cite{balzarotti2008saner,doupe2012enemy}. These have become less effective in present times due to the increasing complexity of web applications, their dependency on downstream technologies, and the limited development and deployment time~\cite{wichers2013owasp}. Worse yet, the attackers, with time on their side, can perform reconnaissance and attack. To address this challenge, we consider a Moving Target Defense (MTD) based approach~\cite{mtd-cn}, which complements the existing vulnerability analysis techniques through a defense-in-depth mechanism.

The MTD based approach dynamically shifts a system over time to increase the uncertainty and complexity for the attackers to perform probing and attacking \cite{zhuang2014towards}, while ensuring that the system is available for legitimate users. As the window of attack opportunities decreases, the effort in finding and successfully executing an attack increases. Moreover, if an attacker succeeds in finding a vulnerability at one point in time, it may not be exploitable at another time because of the moving defense system, making the web application more resilient~\cite{taguinodtoward2015}.

Various aspects that support Moving Target Defense approach such as, using multiple implementation languages, multiple database instances with synchronization, etc. are considered in different layers of web application architecture, along with ways to switch between them. \emph{However, the design of good quality switching strategies itself is left as an open problem}. This is key to effectively leverage various move options---thereby maximizing the complexity for the attacker and minimizing the damage for the defender.

Our aim in this paper is to design effective switching policies for movement in the MTD system that maximize the security of the web application, given the set of components and configurations of the system which can be ``moved around'', while simultaneously considering realistic costs for ``moving them around''. In web applications, the defender (leader) deploys a system up-front. The attacker observes (or follows) the system over time before choosing an attack. These characteristics motivate us to formulate the MTD system as a repeated Bayesian Stackelberg Game (BSG).  For this formulation to be meaningful in real-world applications, we use real world attack data for our model.  We propose a framework to define attacker types for our game and automatically generate attack options for each of them by mining and characterizing Common Vulnerabilities and Exposures (CVEs).  We develop a system that leverages the knowledge in public attack databases and expertise of system administrators for obtaining meaningful game utilities and switching costs respectively.

For computing the movement policy for the defender, we initially expected to be able to use existing solvers developed for  physical security systems \cite{Sinha16a}.  Unfortunately, none of them considered the cost of switching between strategies.  Since this is highly relevant in cyber-security systems,  we had to formulate an optimization problem to consider these costs when generating strategies.

The increased complexity in an MTD systems exacerbates the difficulty of prioritizing vulnerabilities that need to be fixed next. We define this problem formally and propose a preliminary solution. Lastly, we talk about metrics to measure the robustness of switching strategies generated by various models when the uncertainty about attacker types vary in the real world.

In \autoref{sec:rw}, we introduce the reader to the different ways people have tried to address the problem of generating switching strategies for cyber-security systems. We introduce domain terminology related to MTD systems for web-applications in \autoref{sec:3}. In \autoref{sec:4}, we develop the Bayesian Game model for this system defining attacker types, attack classification, rewards generated from security databases and strategy switching costs. To find an effective switching strategy, we propose a solver that maximizes system security while accounting for switching costs in \autoref{sec:5}.  We empirically study the effectiveness and robustness of the strategy generated by our framework in \autoref{sec:6}, comparing it to the state-of-the-art.  We also formulate the problem of identifying critical vulnerabilities and propose a preliminary solution in \autoref{sec:6}.  We conclude the paper in \autoref{sec:7}, highlighting promising research directions.


\section{Related Work}
\label{sec:rw}


Although there exists prior work on the design of switching strategies for MTD systems, most of it is domain specific. Evaluation of these strategies on real-world MTD systems for web applications is scarce. We discuss some of these works, highlighting their limitations in the domain of web applications, thus motivating the need for our solution. Existing efforts describe the use of randomized switching strategies, and show its effectiveness for MTD systems \cite{zhuang2014towards}. We empirically demonstrate that our strategy outperforms this state-of-the-art for web applications, especially when the cost of switching is negligible.


Attacker-defender scenarios have been modeled earlier as stochastic games for attack-surface shifting \cite{surface}. Other works model the MTD problem as a repeated game where the defender uses uniform random strategy with the exception that the same defense configuration is not deployed in two consecutive rounds \cite{winterrose2014adaptive}. This work needs an in-depth analysis of code, which is unrealistic for complex web applications.

Switching strategies for MTD systems based on detection of probes by attackers are presented by \cite{Prakash15EGA}. Unfortunately, an accurate detection of attacks in web applications is difficult, if not impossible. Furthermore, such strategies can lead to a detrimental performance in repeated games if an intelligent attacker biases the system to switch more towards MTD configurations where the attacker attains higher reward. In~\cite{jones2015evaluating}, the MTD system is modeled as a game called PLADD, based on FlipIt games \cite{van2013flipit}. This work assumes that different agents control the server in different game rounds, which is impractical for most cyber-seccurity applications, essentially web applications. These techniques also fail to capture the reconnaissance aspect of the attackers which is shown to be an important aspect in the attack phase \cite{okhravi2014finding}.

In~\cite{gt-dyna}, a game theoretic leader-follower type approach is presented for a dynamic platform defense where the strategies are chosen so as to be diverse, based on statistical analysis rather than being uniformly distributed. They find similarity among different configurations of the MTD system, which is difficult in the domain of web applications. The work fails to consider the uncertainty in the attacker model and the costs for switching.

These aspects of uncertainty in the attacker model and attacker reconnaissance are handled effectively via Bayesian Stackelberg Games (BSG), making it an appropriate choice for modeling the web applications domain. Our modeling could help us leverage the existing solution methods in the physical security domains \cite{Sinha16a} and provide scalable and optimal switching strategies for cyber-security systems.  Unfortunately, these works, to our knowledge, do not consider the cost the defenders incur when asked to switch from a particular strategy to another.  Hence, we propose a solver that maximizes the defender's reward and minimizes the overall cost of switching between web-application configurations. Our solver is essentially an extension of the DOBSS solver \cite{dobss}.  Although there has been furhter development since DOBSS, the more recent solvers for BSGs make additional assumptions about the game structure---either about the action sets of the defender, or the presence of hierarchical structure among attacker types~\cite{amingradient}, which do not hold for the web application domain.

The use of Common Vulnerabilitiy Scoring System (CVSS) for rating attacks is well studied in security \cite{houmb2010quantifying}. We describe this metric later. CVSS provides a strong backbone for obtaining utilities for our game theoretic model. None of the existing works (to our knowledge) talk about the pragmatic aspect of prioritizing vulnerabilites in MTD systems. Also, there does not seem to be any standard metrics to capture the robustness of strategies generated by a model. We address both these issues in the upcoming sections. 
\section{Moving Target Defense for Web Applications}
\label{sec:3}
In this section, we present a brief overview of the web application domain and its functionality which will be useful for understanding the challenges involved in generating solution strategies.


\subsection{Configuration} A configuration set for a web application stack is denoted as $C = C_1 \times C_2 \dots \times C_n$ where there are $n$-technological stacks. Here, $C_i$ denotes the set of technologies that can be used in the $i$-th layer of the application stack. A \textit{valid} configuration $c$ is an $n$-tuple that preserves the system's operational goals.

Consider a web application that has two layers ($n=2$) where the first layer denotes the coding language the web-application was coded in and the second layer denotes the database that stores the data handled by this application. Say, the technologies used in each layer are $C_1=\{$Python, PHP$\}$ and $C_2=\{$MySQL, postgreSQL$\}$.  A valid configuration can be $($PHP, MySQL$)$.  The \textit{diversity} of an MTD system, which is the number of valid configurations, can be 4 (at max) in this case.

\subsection{Attack}
Software security is defined in terms of three characteristics - Confidentiality, Integrity and Availability \cite{mccumber1991information}. In a broad sense, an attack on a web application is defined as an \textit{act} that compromises any of the aforementioned characteristics. The National Vulnerability Database (NVD) is a public directory of known vulnerabilities and exposures affecting all technologies that can be used in a web application. The Common Vulnerabilities and Exploits (CVEs) in this database list vulnerabilities and corresponding attacks that can be used to compromise an application using the affected technology. As each CVE has an exploit associated with it,
we use the terms \textit{vulnerability} and \textit{attack} interchangeably going forward.

\subsection{Switching Strategy}
This is a decision making process for the defender to select the next valid system configuration $c'$ given $c$ as the present system configuration (where both $c,c' \in C$). If $p_c$ represents the probability that $c$ is chosen in a given deployment cycle through randomization, a switching strategy is $f:C\rightarrow p_c$ where $\sum\limits_{c \in C} p_c =1~\forall~p_c\in[0,1]$. To add to the complexity, the cost for switching from a configuration $c$ to another configuration $c'$ can be nontrivial and non-uniform.  Thus, the aim of a good strategy is to maximize the effectiveness of an MTD system while trying to minimize the cost for switching. Present state-of-the-art MTD system for web applications use a uniformly distributed switching strategy ($p_c=1/|C|$) and assume that switching between configurations incur a uniform cost \cite{taguinodtoward2015}.

We now develop a game theoretic system to generate switching strategies for the MTD web application that 1) shows a uniformly distributed switching strategy is sub-optimal and 2) considers the non-negative non-uniform costs of switching between different configurations of an MTD system.

\section{Game Theoretic Modeling}
\label{sec:4}
In this section, we model the setup of MTD systems in as a repeated step Bayesian Game. 

\subsection{Agents and Agent types}
There are ($N=$) two players in our game, a defender and an attacker. The set $\theta_i$ is the set of types for player $i~(=\{1,2\})$. Thus, $\theta_1$ and $\theta_2$ denotes the set of defender and attacker types respectively.  The $j-$th attacker type is represented by $\theta_{2j}$.

When an attacker attacks an application, its beliefs about what (resource/data) is most valuable to the application owner (defender) remains consistent. Thus, we assume that the attacker knows that there is only one type of defender when (s)he attacks a particular web application.  Thus, we have $|\theta_1|=1$.

We consider finite types of attackers. Each attacker type is defined in our model using a 3 tuple,
\[
\theta_{2i}=\langle name, \{(expertise, technologies)\dots\}, probability\rangle
\]%
where the second field is a set of two dimentional values that express an attacker's expertise ($\in [0,10]$) in a technology. The rationale for using values in this range stems from the use of Common Vulnerability Scoring System (CVSS) described later. Lastly, the set of attacker types have a discrete probability distribution associated with it. The probability $P_{\theta_{2j}}$ represents the defender's belief about the attacker type $\theta_{2j}$ attacking their application. Obviously, the probability values of all attacker types sum up to one.
\[
\sum_{ \theta_{2j} \in \theta_{2} } P_{\theta_{2j}} = 1
\]

Note that one can define attacker expertise over a `\textit{category of attacks}' (like \textit{`BufferOverflowAttacks'}) instead of \textit{technology} specific attacks. We feel the latter is more realistic for our domain. This definition captures the aspect that an attacker type can have expertise in a set of \textit{technologies}. Since, these attacker types and the probability distribution over them are application specific, it is defined by a domain expert and taken as an input to our proposed model. For instance, a defender using a no-SQL database in all configurations of his MTD system, assigns zero probability to an `\textit{SQL\_database}' attacker type because none of their attacks can compromise the security of his present system.


The assumption that the input probability distribution over all the attacker types can be accurately specified is a strong one. We later discuss how errors in judgment can affect the effectiveness of a switching strategy and define a measure to capture the robustness of the generated policy in such circumstances.

\subsection{Agent actions}
We define $A_{\theta_i}$ as a finite set of actions available to player $i$. The defender action set, $A_{\theta_1}$ is a switch action to a valid configuration, $c$ of the web application. The maximum number of  actions (or pure strategies) for the defender can ideally be $|C_1| \times |C_2| \dots \times |C_n|$. This might be lower since a technology used in layer $x$ might not be compatible when paired with a technology  used in layer $y~(\neq x)$ rendering that configuration \emph{invalid}.

For the attacker, $A_{\theta_2}$ represents the set of all attacks used by atleast one attacker type. A particular attack $a$ belongs to the set $A_{\theta_2}$ if it affects atleast one of the technologies used in the layers for our web application ($C_1\cup C_2\dots\cup C_n$).

We now define a function $f:(\theta_{2t}, a) \rightarrow \{1,0\}$ for our model. The function implies an attack $a$ is a part of the attacker type $\theta_{2t}$'s arsenal $A_{\theta_{2t}} (\subseteq A_{\theta_2})$ if the value of the function is 1. This function value is based on the similarity between (i) the expertise of the attacker type contrasted with the `exploitability' necessary to execute the attack, and (ii) the attacker's expertise in the technology for which the attack can be used. We provide a concrete definition for the function $f$ after elaborating on what we mean by exploitability of an attack.

For (almost all) CVEs listed in the NVD database, we have a six-dimensional CVSS v2 vector representing two independent scores -- Impact Score (IS) and Exploitability Score (ES).  For an attack action $a$, $ES_a$ ($\in [0,10]$) represents the ease of exploitability (higher is tougher). For each attack, the database also lists a set of technologies it affects, say $T^a$.

Let us consider the set of technologies an attacker type $t$ has expertise in is $T_t$. Now we define the function $f$ as,
\begin{eqnarray}
f(\theta_{2t}, a) &=&
\left\{
	\begin{array}{ll}
		1, & \textit{iff } T_t \cap T^a \neq \phi~\land ~ES_a \leq \textit{expertise}_t\\
		0 & \textit{otherwise}
	\end{array} \nonumber
\right.
\end{eqnarray}

\subsection{Reward values for the Game}
Now that we have attack sets for each attacker type, the general reward structure for the proposed game is defined as follows:
%
\begin{eqnarray}
R_{{a,\theta_{2i}},c}^A =
\left\{
	\begin{array}{ll}
		+x_a & \mbox{if }a \subset \upsilon(c) \\
		-y_a & \mbox{if } a~\mbox{can be detected or } a \subset c' \\
		0 & ~\mbox{otherwise}
	\end{array} \nonumber
\right. \\
R_{{a,\theta_{2i}},c}^D =
\left\{
	\begin{array}{ll}
		-x_d & \mbox{if }a \subset \upsilon(c) \\
		+y_d & \mbox{if } a~\mbox{can be detected or }a \subset c' \\
		0 & ~\mbox{otherwise}
	\end{array} \nonumber
\right.
\end{eqnarray}
where {\small $R_{a,\theta_{2i},c}^A$} and {\small $R_{a,\theta_{2i},c}^D$} are the rewards for the attacker type and the defender respectively, when the attacker type $\theta_{2i}$ uses an attack action $a$ against a configuration $c~(\in C)$. The function $\upsilon(c)$ represents the set of security vulnerabilities (CVEs) that configuration $c$ has. Also, $c'$ refers to a honey-net configuration.  A honey-net is a configuration setup with intentional vulnerabilities to invite attackers for catching (or observing) them.

Note that the reward values when a attacker does not attack (NO-OP action), is zero. Moreover, a defender gets zero reward for successfully defending a system. We reward him positively only if he/she is able to reveal some more information or catch the attacker without impacting operation requirements for the non-malicious users (or using honey-nets). He gets a negative reward if an attacker successfully exploits his(/her) system.

To obtain reward values for the variables $x_a, y_a, x_d$ and $y_d$, we make use of CVSSv2 metric.
This metric provides the Impact (IS) and Exploitability Scores (ES), stated above, which are combined to calculate a third score called Base Score (BS) \cite{mell2007cvss}. Using these, we now define the following:
{\small\begin{eqnarray}
x_d &=& -1 * IS  \nonumber \\
x_a &=& BS  \nonumber
\end{eqnarray}}%
\indent Note that BS considers both the impact and the exploitability. When the IS for two attacks are the same, the one that is easier to exploit gets the attacker a higher reward value. The ease of an attack can be interpreted in terms of the resource and effort spent by an attacker for an attack \emph{Vs.} the reward (s)he attains by harming the defender. Although the robustness of our framework provides provisions for having $y_d$ and $y_a$, detecting attacks on a deployed system or setting up honey-nets is still in its nascent stages.
Hence, there are no actions where values of  $y_d$ or $y_a$ are required in our present application.

Before we move on, we describe briefly what security dimensions the independent scores (IS and ES) are actually trying to capture in the context of a real world software system.  For this purpose, we first define the 6 independent values that generate these scores.
\begin{itemize}
    \item \textbf{Access Vector (AV)} is dependent on the amount of access an attacker needs to exploit a vulnerability.  Thus, an attack that needs physical access to a system will have lower score than one that can be exploited over the Internet by any machine.
    \item \textbf{Access Complexity (AC)} represents the complexity of exploiting an attack.  A buffer overflow attack on an Internet service is less complex than an e-mail client vulnerability in which a user has perform attachment downloads followed by executing it and hence has lower AC value.
    \item \textbf{Authentication (Au)} level required to execute the attack.  For example, if no sign-up account is required to exploit the system, this value is high.  In contrast, if one needs multiple accounts to exploit the vulnerability, the value is low.
    \item \textbf{Confidentiality Impact (C)} scores are low if only some (non-relevant) information gets leaked.  Highest impact occurs when say, the entire database is compromised if the vulnerability is successfully exploited.
    \item \textbf{Integrity Impact (I)} refers to the attacker's power to modify files or behaviour of a system if he executes the exploit successfully. The more the power-- say the attacker is able to change code or remove arbitrary files in the system-- the higher this value.
    \item \textbf{Availability Impact (A)} represents the power of a successful exploit to bring down the availability of a system.  A successful Denial of Service (DoS) that brings down an application server, will have high impact.
\end{itemize}%
From these values, one can obtain the two independent scores using the following formulas,
\begin{eqnarray}
ES &=& 20*(AV)*(AC)*(Au)  \nonumber \\
IS &=& -10.41*(1-(1-C)(1-I)(1-A)) \nonumber
\end{eqnarray}%
A rigorous treatment of assigning these values can be found in \cite{mell2007cvss}.  The CVSS values are generated by security experts across the globe and the database is updated every single day.

Our model takes a time range as input. It then parses all the CVEs ($a$) from the NVD in that time range to finally filter out the ones that can affect atleast one of the configurations in our system ($a \subset \upsilon(c_i)$). Note that old CVEs are irrelevant for generating attack sets for a relatively new MTD system as they either have no effect on the updated versions of the technologies they can affect or have popular solutions to prevent them while developing the application. For our application, we obtain this input range from our security experts.

\subsection{Switching Cost}
The switching costs can be represented by a $K^{n\times n}$ matrix where the $n$ rows (and columns) denote the $n$ system configurations.  The cell $K_{ij}$ denotes the cost of switching when the defender moves from configuration $i$ to configuration $j$.  As mentioned earlier, the values in $K$ are all non-negative.  Our security experts, who have written the code to automatically move from one configuration to another, hand code these values in each cell of the martix.  We provide some guidance in choosing these values here and give a concrete example on how we selected these for our application later.

If there is no common technology between configurations $c$ and $c'$ involved in a switch operation, the cost will be large. Also, switching technologies in a specific layer may incur more cost than switching technologies in other layers. In the developed MTD system, we find that switching between databases incur large costs because the structure of the data needs to be changed for shifting, and the time required to copy huge amounts of data from one database to another must also be accounted for.

The matrix $K$ for our system turns out to be symmetric, i.e. $K_{ij} = K_{ji} ~\forall~ i,j\in\{1,\dots n\}$.  Also, $K_{ii}=0$, which implies that there is no cost if no configuration switch occurs.  Note that although our security experts think this is the structure of rewards for the developed system, the modelling is generic enough to allow for asymmetric costs.  Lastly, we choose the values of $K_{ij}$ in the range $[0, 10]$.  The reason for this upper bound becomes clear in the upcoming section.
\section{Switching Strategy Generation}
\label{sec:5}
In this section, we first introduce the notion of Stackelberg Equilibrium for our security game, that gives us a defender strategy that maximizes his reward (and thus the security of the system).  We briefly talk of optimization methods, relevant to our domain, that can produce this.  Finally, we incorporate the costs of switching into the objective function and propose our solver.
\subsection{Stackelberg Equilibrium}
The strategy generated for the designed game needs to capture the reconnaissance aspect. Note that the game starts only after the defender has deployed the web application, acting as a leader. This now becomes a repeated game in which an attacker can observe a finite number of switch moves and probabilistically learn the switching strategy (since $|C| \ll \infty$) of the defender. Thus, the defender has to select a strategy that maximizes his reward in this game, given that the attacker knows his strategy. This is exactly the problem of finding the Stackelberg Equilibrium in a Bayesian Game \cite{von2010market}.  The resulting mixed strategy is the switching strategy for the defender in our MTD system.  Unfortunately, this problem becomes \textit{NP-hard} in our case because of multiple attacker types \cite{bsg}.

Before we find a \textit{strong} Stackelberg Equilibrium for our proposed game, we state a couple of well founded assumptions we make.  Firstly, an attacker chooses a pure strategy, i.e., a single attack action that maximizes his reward value.
This assumption is popular in prior work on security games because for every mixed strategy for the attacker, there is always a pure strategy in support for it~\cite{armor}. Secondly, we assume that the pure strategy of an attacker type is not influenced by the strategy of other attacker types.  This is not limiting for our web application domain since an attacker type's attack selection is independent of the attack action chosen by another type.

To solve for the optimal mixed strategy, one can use the Decomposed Optimal Bayesian Stackelberg Solver (DOBSS)~\cite{dobss}. This optimizes the expected reward of the defender over all possible mixed strategies for the defender ($\vec{x}$), and pure strategies for each attacker type ($\vec{n}^{\theta_{2i}}$) given the attacker type uncertainty ($\vec{P}_{\theta_{2i}}$).  We now define the objective function of the Mixed Integer Quadratic Program (MIQP) in \autoref{eqn}.
\begin{equation}
\max_{x,n,v} \sum\limits_{c\in C}^{} \sum\limits_{\theta_{2i} \in \theta_{2}}^{} \sum\limits_{a \in A_{\theta_{2i}}}^{} P_{\theta_{2i}} R_{a,\theta_{2i},c}^D~x_c n_a^{\theta_{2i}} \label{eqn}
\end{equation}%
Notice that this does not consider that switching costs between defender strategies.  Essentially, this means the formulation assumes that switching costs are uniform. Before we address this limitation in the upcoming subsection, we take a little digression.

For our scenario, we have many attack actions.  Thus, we observe that solving the MIQP version is more efficient (in computation time and memory usage) than solving the Mixed Integer Linear Program (MILP) version of the DOBSS. This can be attributed to the fact that the MILP formulation results in an increase in the dimensions of the solution space. Theoretically, the MIQP solves for $|C|+\sum_{\theta_{2i} \in \theta_2}\sum_{a_j \in A_{\theta_{2i}}}$ $|a_j|$ variables where as the MILP solves for $|C|*\sum_{\theta_{2i} \in \theta_2}\sum_{a_j\in A_{\theta_{2i}}}$ $|a_j|$ variables.

\subsection{Incorporating Switching Costs}

As defined in the last section, the cost for switching from a configuration $i$ to a configuration $j$ can be represented as $K_{ij}$.  The probability the system is in configuration $i$ and then switches to configuration $j$ is $x_i\cdot x_j$.  Thus, the cost incurred by the defender for a switch action from $i$ to $j$ is $K_{ij}\cdot x_i \cdot x_j$.  The expected cost for any switch action is $\sum\limits_{i\in C} \sum\limits_{j\in C} K_{ij} \cdot x_i \cdot x_j$.

To account for cost, we can subtract this from the objective function of \autoref{eqn} with a cost-accountability factor $\alpha~(\geq0)$ to obtain
\[
\max_{x,n,v} \sum\limits_{c\in C}^{} \sum\limits_{\theta_{2i} \in \theta_{2}}^{} \sum\limits_{a \in A_{\theta_{2i}}}^{} P_{\theta_{2i}} R_{a,\theta_{2i},c}^D~x_c n_a^{\theta_{2i}} - \alpha \cdot \sum_{i\in C} \sum_{j\in C} K_{ij} \cdot x_i \cdot x_j
\]%
\noindent Unfortunately, this results in a Bilinear Mixed Integer Programming problem, which is not convex.  To ameliorate this problem, we now introduce new variables $w_{ij}$ that essentially represent an approximate value of $x_i \cdot x_j$.  We first use the piecewise linear McCormick envelopes to design a convex function using these $w_{ij}$-s that estimates a good solution to this problem \cite{wicaksono2008piecewise}.  Along with these constrains, we introduce further constrains which we describe after introducing the final MIQP convex optimization problem as follows,
{\small
\begin{equation}
\max_{x,n,v} \sum\limits_{c\in C}^{} \sum\limits_{\theta_{2i} \in \theta_{2}}^{} \sum\limits_{a \in A_{\theta_{2i}}}^{} P_{\theta_{2i}} R_{a,\theta_{2i},c}^D~x_c n_a^{\theta_{2i}} ~-~ \alpha \cdot \sum_{i\in C} \sum_{j\in C} K_{ij} w_{ij} \label{eqn:2}
\end{equation}%
}%
\begin{eqnarray}
s.t.~~~ \sum\limits_{c\in C}^{} x_c &=& 1  \\
\sum\limits_{a\in A_{\theta_{2i}}}^{} n_a^{\theta_{2i}} &=& 1  \\
0 \leq v^{\theta_{2i}} - \sum\limits_{c\in C} R_{a,\theta_{2i},c}^A x_c  &\leq& (1-n_a^{\theta_{2i}})M \\
w_{ij} &\geq& 0 ~\forall~ i,j \\
w_{ij} &\leq& x_i ~\forall~ i,j \\
w_{ij} &\leq& x_j ~\forall~ i,j \\
\sum\limits_{j \in C} \sum\limits_{i \in C} w_{ij} &=& 1 ~\forall~ i,j\\
\sum\limits_{j \in C} w_{ij} &=& x_i ~\forall~ i\\
\sum\limits_{i \in C} w_{ij} &=& x_j ~\forall~ j
\end{eqnarray}
\begin{equation*} 
    x_c \in [0\dots 1],~ n_a^{\theta_{2i}} \in \{0,1\},~ v^{\theta_{2i}} \in \mathcal{R}
\end{equation*}
\[ \forall ~c \in C \nonumber,~\theta_{2i} \in \theta_2,~a \in A_{\theta_{2i}} \]
where $M$ is a large positive number. $\vec{n}^{\theta_{2i}}$ and $v^{\theta_{2i}}$
give the pure strategy and its corresponding reward for the attacker type $\theta_{2i}$ respectively, and $\vec{x}$ gives the mixed switching strategy for the defender.  (5) solves the dual problem of maximizing rewards for each attacker type ($v^{\theta_{2i}}$) given the defender's strategy. This ensures that attackers always select the best attack action.
The constrains (6), (7) and (8) represent the McCormick envelope that provides lower and upper bounds on each $w_{ij}$.
Since we consider all possible switches, $\sum\limits_{j \in C} \sum\limits_{i \in C} x_{i} \cdot x_j = 1$.  This is enforced by constrain (9).
Lastly, for each $i$, $\sum\limits_{j \in C} x_i \cdot x_j = x_i \cdot (\sum\limits_{j \in C} x_j) = x_i$.  This is represented by the constrains (10) and (11).

If we now allow the maximum cost of switching to be $10$, we can see that the values for the cost is comparable in magnitude to the value of the defenders rewards.  This helps us to provide a semantic meaning for the cost-accountability factor, $\alpha$.

The first term in the objective function seeks to maximize the defender's reward, which in turn maximizes the security of the web application.  The second term on the other hand, seeks to reduce the expected cost of the switching actions.  Thus, $\alpha$ represents how much importance is given to the cost of switching \textit{Vs.} the level of security desired.  Consider the extreme cases, when $\alpha = 0$, we are producing the most secure strategy (which is the Stackelberg Equilibrium) and considering switching between configurations incur zero costs.  This is the sub-set of solutions that are developed mostly for physical security systems.  In contrast to that, when $\alpha = 1$, we are saying that we consider switching costs as important as the security of our MTD application.

Choosing the correct value of $\alpha$ is not trivial and often dependent on the specific web-application.  For example, if a banking system someday seeks to operate on a MTD system, we \textit{hope} it puts more weight on security than switching costs, selecting low $\alpha$ values.  To provide a sense to the reader, we later show in the experimental section, how strategies and reward values are effected with changing alpha values.

\section{Empirical Evaluation}
\label{sec:6}

\begin{figure}[t]
    \includegraphics[width=3in]{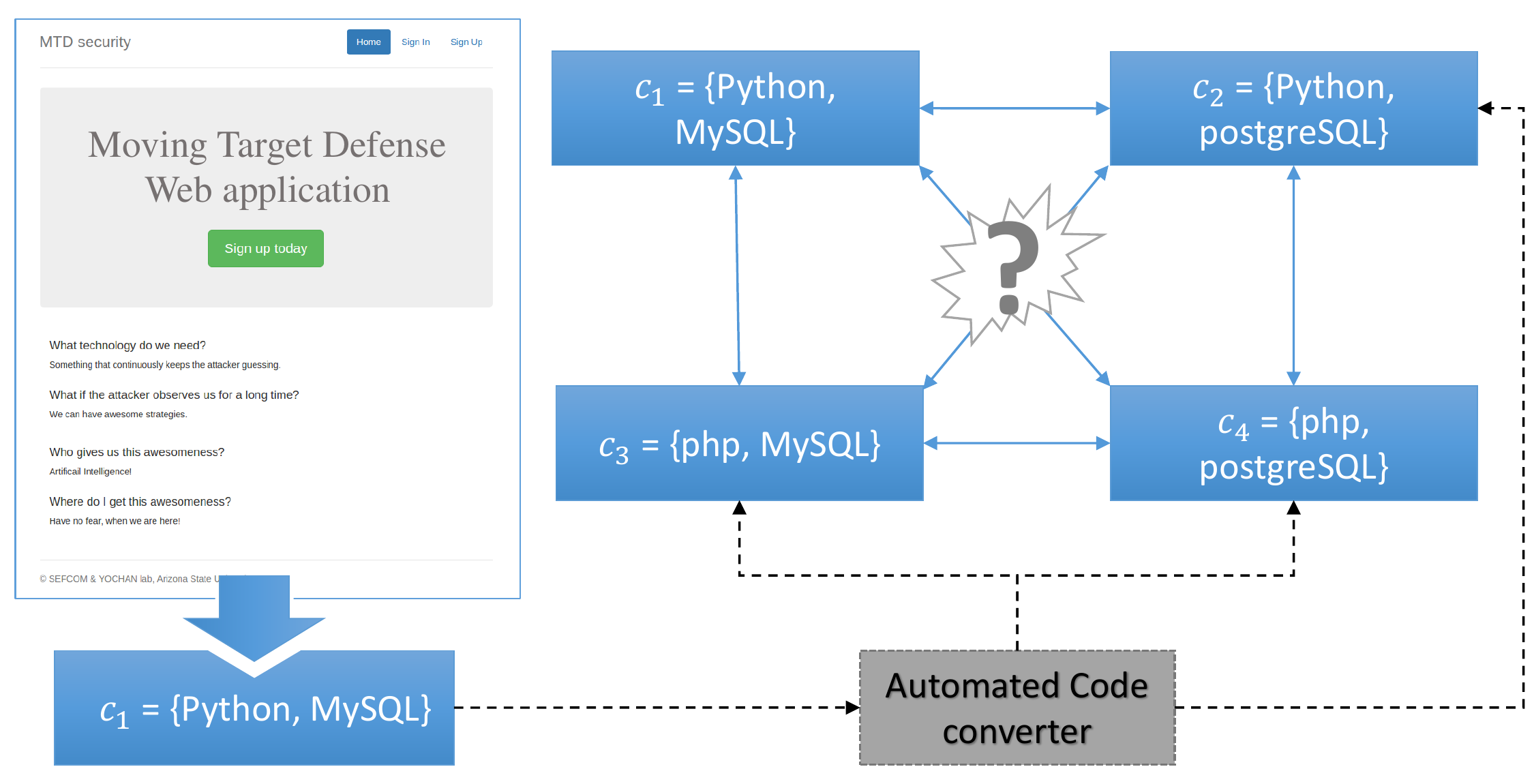}
    \caption{A moving target defense web application system}
    \label{web-app-struct}
\end{figure}%

The goal of this section is to answer three key questions.  Firstly, does our proposed Bayesian Stackelberg Game (BSG) model generate better strategies that the state-of-the-art?  Secondly, can we effectively compute the set of critical vulnerabilities? Lastly, who are the sensitive attacker types and how robust is our model?

\subsection{Test Bed Description}

To answer the questions mentioned above, we develop a real world MTD web application (shown in \autoref{web-app-struct}) with 2 layers.  The key idea of applying MTD to web applications requires you to have several versions of the same system, each written in either a different language, using a different database, etc.

This diversity is not ubiquitous in legacy web applications, due to cost, time, and resources required to build several versions of the same web application.  To aid this, we developed a framework to automatically generate the diversity necessary for this web application. The current prototype is able to convert a web application coded in Python to an equivalent one coded in PHP, and vice versa, as well as a web application using a MySQL database to an identical version that uses PostgreSQL, and vice versa. In the future, as more and more variations are developed, the set of defender's actions will increase.

The present set of valid configurations for our system is $C=\{($PHP, MySQL$)$, $($Python, MySQL$)$, $($PHP, postgreSQL$)$, $($Python, postgreSQL $)\}$.  The costs for switching between configurations is shown in \autoref{table:2}.  These cost values generated by our system administrators are based on the following considerations:
\begin{itemize}
    \item Switching between different languages while keeping the same database dialect incurs minimal cost - workload is primarily done  is primarily on rerouting to the correct server with the source language
    \item Switching between different database dialects while keeping the same language incurs a slightly higher cost due to the conversion required for the database structure and its contents.  One also has to account for copying large amounts of data to the database used in the current system configuration.
    \item Switching between different database dialects AND different languages incur the most cost due to the combination of the costs of conversion for the database as well as the penalty for rerouting to the correct server with the source language.
\end{itemize}

The attacker types along with the attack action set size are defined in \autoref{table:1}.  We mined the NVD for obtaining CVE data from January, 2013 to August, 2016 to generate these attack sets. If the stakes of getting caught are too high for an attacker type given an MTD system, he/she may choose not to attack. Hence, we have a \textit{NO-OP} action for each attacker type.

\begin{table}[t]
\small
\centering
 \begin{tabular}{||p{1.4cm} p{1.1cm} p{1.1cm} p{1.2cm} p{1.2cm}||} 
 \hline
  & PHP, MySQL & Python, MySQL & PHP, postgreSQL & Python, postgreSQL\\ [0.5ex] 
 \hline\hline
 \rowcolor{lightgray} PHP, MySQL & 0 & 2 & 6 & 10\\
 Python, MySQL & 2 & 0 & 9 & 5\\
 \rowcolor{lightgray} PHP, postgreSQL & 6 & 9 & 0 & \cellcolor{yellow!25}2\\
 Python, postgreSQL & 10 & 5 & \cellcolor{yellow!25}2 & 0\\
 \hline
 
 \end{tabular}
 \caption{Swithing costs for our system}
 \label{table:2}
\end{table}

The optimization problems for the experiments were solved using Gurobi on an Intel Xeon $E5~2643v3@ 3.40$GHz machine with $6$ cores and $64$GB of RAM.

\begin{table}[t]
\centering
 \begin{tabular}{||p{2cm} p{2cm} p{1.1cm} p{0.7cm} ||} 
 \hline
 Name & (Technologies, Expertise) & Prob. & $|A_{\theta_{2i}}|$\\ [0.5ex] 
 \hline\hline
 \rowcolor{lightgray} Script Kiddie (SK) & (PHP,4), (MySQL,4) & 0.15 & 34\\
 Database Hacker (DH) & (MySQL,10), (postgreSQL,8) & 0.35  & 269\\
 \rowcolor{lightgray} Mainstream Hacker (MH) & (Python,4), (PHP,6), (MySQL,5) & 0.5 & 48\\
 \hline
 \end{tabular}
 \caption{Attacker types and attack action counts}
 \label{table:1}
\end{table}

\subsection{Strategy Evaluation}

\begin{figure}[t]
\centering
\resizebox{0.45\textwidth}{!}{
    \begin{tikzpicture}[]
        \begin{axis}[
        xticklabel style={anchor=near xticklabel},
        xlabel={$\alpha$},
        ylabel={Obj},
        legend style={
            cells={anchor=west},
            draw=none, fill=none, 
            font=\scriptsize,
            legend pos= north east,
        },
        ]
        \pgfplotsset{height=5cm}
        \addplot table[x=Alpha,y=BSG] {obj.dat};
        \addplot table[x=Alpha,y=URS] {obj.dat};
        \addlegendentry{$BSG$}
        \addlegendentry{$URS$}
    \end{axis}
    \end{tikzpicture}
}
\caption{Objective function values for Uniform Random Strategy \textit{Vs.} Bayesian Stackelberg Game with switching costs as $\alpha$ varies from $0$ to $1$.}
\label{fig:2}
\end{figure}
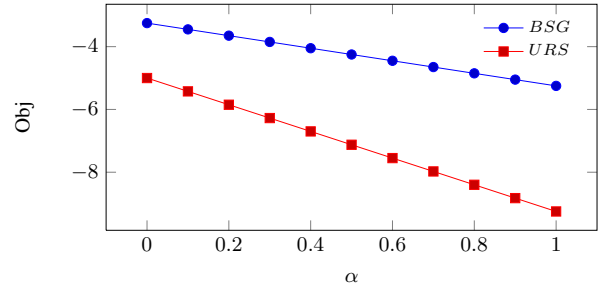

We evaluate our method using Bayesian Stackelberg Games on our real life web application against the Uniform Random Strategy (URS), which is the state-of-the-art in such systems \cite{taguinodtoward2015}.  We plot the values of the objective function in \autoref{eqn:2} for both the strategies as $\alpha$ varies from $0$ to $1$.  For URS, we use the exact values of $w_{ij} = 0.25*0.25=0.0625 ~\forall~ i,j$.  The plot is shown in \autoref{fig:2}.  Both are straight lines because although the value of $\alpha$ changes, the strategy for URS is same (by definition) and the one generated by BSG also remains the same.  The latter case came as a surprise to us initially.  On further investigation, we noticed that in the formulated game for our web-application, the Stackelberg Equilibrium for our application (luckily) coincides with the least switching cost strategy.



These attacker and defender strategies is shown in \autoref{table:3} alongwith the value of the defender's reward (i.e. the first term in the objective function in \autoref{eqn:2}).  Notice that, not only is the mixed strategy generated by BSG more secure than URS, it leverages fewer configurations than all valid configurations $|C|=4$ the system has to offer.  This result is in unison with previous work in cyber-security which show that having many configurations does not necessary imply that all of them have to be used for providing the best security \cite{gt-dyna}.

\begin{table}[t]
\small
\centering
 \begin{tabular}{||p{0.8cm} p{2.2cm} p{1.2cm} p{2.2cm}||} 
 \hline
 Method & Mixed Strategy & Defender's Reward & Attack sets (SK, DH, MH)\\ [0.5ex] 
 \hline\hline
 \rowcolor{gray!25}URS & (0.25, 0.25, 0.25, 0.25) & -5 & CVE-2016-3477, CVE-2015-3144, CVE-2016-3477 \\
 BSG & (0, 0, 0.5, 0.5) & \cellcolor{green!25}-3.25 &CVE-2014-0185, CVE-2014-0067, CVE-2014-0185 \\
 \hline
 
 \end{tabular}
 \caption{Comparison between the strategies generated by Uniform Random Strategy (URS) \emph{Vs.} Bayesian Stackelberg Game (BSG)}
 \label{table:3}
\end{table}

\subsubsection{Studying the effect of $\alpha$-values}

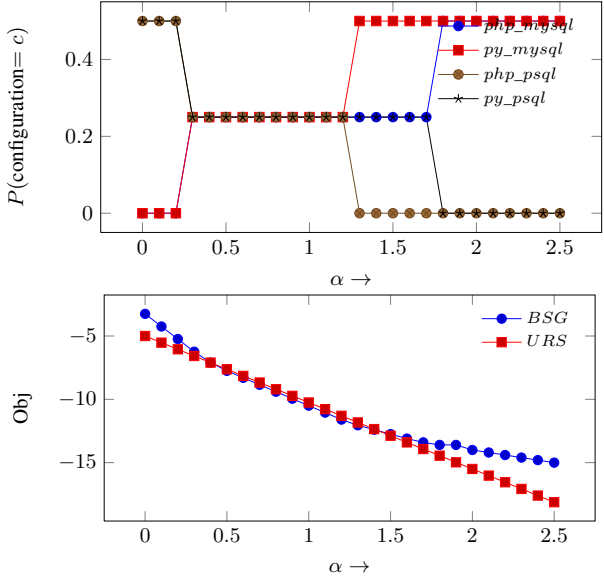
\begin{figure}[t]

\centering
\resizebox{0.46\textwidth}{!}{
    \begin{tikzpicture}[]
        \begin{axis}[
        xticklabel style={anchor=near xticklabel},
        xlabel={$\alpha \rightarrow$},
        ylabel={$P($configuration$=c)$},
        legend style={
            cells={anchor=west},
            draw=none, fill=none, 
            font=\scriptsize,
            legend pos= north east,
        },
        ]
        \pgfplotsset{height=5cm}
        \addplot table[x=Alpha,y=php_mysql] {stra.dat};
        \addplot table[x=Alpha,y=py_mysql] {stra.dat};
        \addplot table[x=Alpha,y=php_psql] {stra.dat};
        \addplot table[x=Alpha,y=py_psql] {stra.dat};
        \addlegendentry{$php\_mysql$}
        \addlegendentry{$py\_mysql$}
        \addlegendentry{$php\_psql$}
        \addlegendentry{$py\_psql$}
    \end{axis}
    \end{tikzpicture}
}

\centering
\resizebox{0.45\textwidth}{!}{
    \begin{tikzpicture}[]
        \begin{axis}[
        xticklabel style={anchor=near xticklabel},
        xlabel={$\alpha \rightarrow$},
        ylabel={Obj},
        legend style={
            cells={anchor=west},
            draw=none, fill=none, 
            font=\scriptsize,
            legend pos= north east,
        },
        ]
        \pgfplotsset{height=5cm}
        \addplot table[x=Alpha,y=BSG] {obj_alpha.dat};
        \addplot table[x=Alpha,y=URS] {obj_alpha.dat};
        \addlegendentry{$BSG$}
        \addlegendentry{$URS$}
    \end{axis}
    \end{tikzpicture}
}
\caption{\textit{Top:} Showcases the change in probabilities associated with a particular configuration.~~\textit{Bottom:} Objective function values for Uniform Random Strategy \textit{Vs.} Bayesian Stackelberg Game with switching costs as $\alpha$ varies from $0$ to $2.5$ when the cost of switching are as showcased in \autoref{table:2} with the values in the yellow boxes being 10.}
\label{fig:6}
\end{figure}

To empirically show that our solver is actually considering costs of switching, we change the value for switching from $($PHP, postgreSQL $)$ to $($Python, postgreSQL$)$ and vice-versa from $2$ (yellow boxes in \autoref{table:2}) to $10$. We plot this scenario in \autoref{fig:6}. As soon as $\alpha \geq 0.4$, the BSG generates $(0.25, 0.25, 0.25, 0.25, 0.25)$ (which is URS) as the most optimal strategy.  After analysis, we note that this happens because the most powerful attack actions in the arsenal of the attacker types are for the systems $($PHP, MySQL$)$ and $($Python, MySQL$)$.  When, one does not prioritize switching costs ($\alpha \in \{0, 0.1, 0.2, 0.3\}$), the system keeps switching between the more secure configurations nullifying the good attacks of the attackers.  As switching costs start to get more significant ($\alpha \in \{0.4, 0.5, \dots 1.2\}$), the objective function value reduces if it sticks to the stronger configurations since switching costs are now high for these.  It switches to the URS in this case.  Beyond that, it switches to the strategy $(0.25, 0.5, 0, 0.25)$ as $\alpha$ keeps on increasing. When $\alpha$ becomes close to $2$, it completely ignores the security of the system and tries to minimize the switching cost by proposing the strategy $(0.5,0.5,$ $0,0)$ as the cost for switching between $($PHP, MySQL$)$ and $($Python, MySQL$)$ is the least ($=2$).

In the bottom of \autoref{fig:6}, we showcase the change in the values of objective function.  At the start, the BSG generates a better strategy when compared to URS.  When the BSG strategy becomes the same as the URS (for $0.4 \leq \alpha \leq 1.2$), we observe that the objective function value for BSG is lower than URS.  This is not surprising since BSG is merely trying to estimate the value $x_i \cdot x_j$ with the variables $w_{ij}$, whereas URS is using the exact value.  As we increase $\alpha$ further, we are essentially discouraging an MTD system, since now the cost of switching has become so high, whereas naive URS pays no heed to this.



\subsection{Identifying Critical Vulnerabilities}

In real-world development teams, it is impossible to solve all the vulnerabilities, especially in a system with so many technologies.  In current software systems, given a set of vulnerabilities, a challenging question often asked is which vulnerabilities should one fix to improve the security?

For an MTD system, this becomes a tough problem since the defender needs to reason about multiple attacker types-- their probabilities and attack actions.  For a given $k$, the set of $k$ vulnerabilities, which on being fixed, result in the highest gain in defender strategy, is termed as the $k$ critical vulnerability set ($k-$CV).

To address this problem, we remove each $k$-sized attack set from the set of all attacks ($A_2' = A_2 \setminus D~\forall~D\subset A_2~\&~|D| = k$) and evaluate the objective function (\autoref{eqn:2}). The sets $A_2'$ that yield the highest objective values, provide the vulnerabilities $D$ that should be fixed to improve the defender's system.

We tried to study this complicated behaviour for some toy examples before applying it to our application.  An interesting phenomenon we noticed was that a $k$-set critical vulnerabilities ($k-$CV) is not always a subset of the $(k+1)-$CV.  Suppose we want to find $3$ vulnerabilities that we want to fix.  Since it is not just a super-set of the $2$-CV, we need to solve this problem from the scratch with $k=3$.  Hence, there is going to be combinatorial explosion here. As the value of $k$ increases, we end up solving $|A_2'| = {|A_2| \choose {k}}$ MIQP problems to identify the $k-$CVs.

\subsubsection{Finding Critical Vulnerabilities in the Developed System}
For our system, we start with $k=1$, we increase number of critical vulnerabilities to be found by $1$ at each step.  The result remains the same for $\alpha \in [0,1]$ for our system.  We do not play around with $\alpha$ beyond this, mostly because this would be unrealistic for any practical application.  Unfortunately, the brute force approach and the scalability of algorithms for solving normal extensive form BSGs proves to be a key limitation.

{\small
\begin{table}[t]
\centering
 \begin{tabular}{||p{0.4cm} p{3cm} p{1.7cm} p{1.5cm}||}
 \hline
 $k$ & CV sets & Objective Value & CPU Time\\ [0.5ex] 
 \hline\hline
 1 & \{(CVE-2014-0185)\} & -2.435 & 3m15s\\
 2 & \{(CVE-2014-0185, CVE-2015-5652)\} & -1.973 & 421m27s\\
 \hline
 
 \end{tabular}
 \caption{Most critical vulnerability in the MTD system and the time required to generate it.}
 \label{table:4}
\end{table}
}%

This is not a surprise since the total number of unique CVEs spread out among the attackers is $287$.  When $k=3$, we end up solving $287 \choose 3$ optimization problems, which fails to scale in both time and memory.  Thus, we only show critical vulnerabilities identified up to $k=2$ (in \autoref{table:4}) using $\alpha = 0.2$.

At present, we are trying to develop a single MIQP formulation that tries to approximately generate the $k$-CV set. To reduce the combinatorial explosion, we plan to use switch variables that can turn attack actions on and off. This comes at the cost of increasing the number of variables in the formulated optimization problem.

\subsection{Model Robustness \& Attacker\\Type Sensitivity}

It is often the case that a web application administrator (defender) cannot accurately specify the probability for a particular attacker type. In this section, we see how this uncertainty affects the optimal rewards generated by the system. We provide a notion for determining sensitive attacker types and measuring the robustness of a switching strategy.

For each attacker type $i$, we vary the probability $P_{\theta_{2i}}$ by $\pm x\%$ $(P_{\theta_{2i}}^{new}=P_{\theta_{2i}}(1\pm \frac{x}{100})$) where $x$ is the \emph{sensitivity factor}, which can be varied from a low value to a high value as needed. Note that now $p=P_{\theta_{2i}}\times\frac{x}{100}$ needs to be adjusted or distributed amongst the probabilities of the remaining attacker types. To make sure that this distribution is done such that the sensitivity of attacker $i$ actually stands out, we propose to distribute $p$ amongst the other attacker types using a weighted model as per their existing probabilities as shown below.
For attacker $j$ ($\neq i$), its new probability would be:
\begin{equation}
\begin{array}{l}
 P_{\theta_{2j}}^{new}=P_{\theta_{2j}}(1\mp \frac{p}{\sum\limits_{k(\neq i)} P_{\theta_{2k}}})
 \end{array}
\end{equation}%
When $x\%$ is subtracted from the probability $\vec{P}_{\theta_{2i}}$, then the sign in the above equation becomes positive, and vice-versa.

We now formally define the loss in reward to the defender as the probability distribution over the attacker types change. 
Let $R_o$ be the overall reward for the defender when he uses the mixed strategy for the assumed (and possibly incorrect) model of attacker type uncertainty ($\vec{P}_{\theta_2}$) on the true model ($\vec{P}_{\theta_2}^{new}$).  Let $R_n$ be the defender's optimal reward value for the true model.
We compute the \emph{Normalized Loss in Rewards (NLR)} for the defender's strategy as follows:
\begin{equation}
\begin{array}{l}
    \emph{NLR} =\frac{R_n - R_o}{R_n}
 \end{array}
 \label{eqn:nlr}
\end{equation}
Note that \emph{NLR} values are $\geq 0$. Higher values of \emph{NLR} represent more sensitive attacker types.  Inaccurate probability estimates for the sensitive attackers can be detrimental to the security of our application. Note that lower \emph{NLR} values indicate that a generated strategy is more robust. 
%

\subsubsection{Evaluation Based on the Developed System}
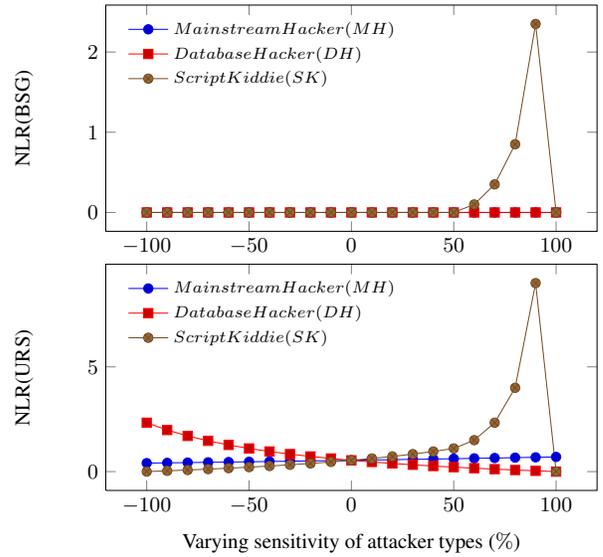
\begin{figure}[t]
\centering
\resizebox{0.45\textwidth}{!}{
    \begin{tikzpicture}[]
        \begin{axis}[
        xticklabel style={anchor=near xticklabel},
        ylabel={NLR(BSG)},
        legend style={
            cells={anchor=west},
            draw=none, fill=none, 
            font=\scriptsize,
            legend pos= north west,
        },
        ]
        \pgfplotsset{height=5cm}
        \addplot table[x=Variations,y=MH] {sensitivity.dat};
        \addplot table[x=Variations,y=DH] {sensitivity.dat};
        \addplot table[x=Variations,y=SK] {sensitivity.dat};
        \addlegendentry{$Mainstream Hacker (MH)$}
        \addlegendentry{$Database Hacker (DH)$}
        \addlegendentry{$Script Kiddie (SK)$}
    \end{axis}
    \end{tikzpicture}
}
\resizebox{0.45\textwidth}{!}{
    \begin{tikzpicture}[]
        \begin{axis}[
        xticklabel style={anchor=near xticklabel},
        xlabel={Varying sensitivity of attacker types ($\%$)},
        ylabel={NLR(URS)},
        legend style={
            cells={anchor=west},
            draw=none, fill=none, 
            font=\scriptsize,
            legend pos= north west,
        },
        ]
        \pgfplotsset{height=5cm}
        \addplot table[x=Variations,y=MH] {URS_sensitivity.dat};
        \addplot table[x=Variations,y=DH] {URS_sensitivity.dat};
        \addplot table[x=Variations,y=SK] {URS_sensitivity.dat};
        \addlegendentry{$Mainstream Hacker (MH)$}
        \addlegendentry{$Database Hacker (DH)$}
        \addlegendentry{$Script Kiddie (SK)$}
    \end{axis}
    \end{tikzpicture}
}
\caption{NLR values for BSG and URS genereated strategies when attacker types probabilities vary in $[-100\%,100\%]$.}
\label{fig:4}
\end{figure}

We compute the attacker sensitivity for our system varying the probability of each attacker type from $-100\%$ to $+100\%$ (of its modeled probability) with $10\%$ step sizes. We plot the results in ~\autoref{fig:4} using \autoref{eqn:nlr}. The Mainstream and Database hacker (MH \& DH) are the least sensitive attacker types.  The \emph{NLR} values for both these attackers are $0$.  This is the case since the real world attack action used by these types remain the same even when their probabilities change.  On the other hand, if the probability associated with the Script Kiddie (SK) is underestimated in our model, we see that the strategies deviate substantially from the optimal.

For our experiments in this section, we use $\alpha=0.2$.  The max \emph{NLR} for our BSG strategy is $2.35$ \emph{Vs.} $9$ for URS.  The average of the 60 \emph{NLR} values is $0.061$ for BSG and $0.88$ for URS. These values indicate our model is more robust to variance in attacker type uncertainty than the present state-of-the-art.
\section{Conclusions and Future Work}
\label{sec:7}
In this paper, we propose a method to generate a switching strategy for real-world web application based on the Moving Target Defense (MTD) architecture. To find an effective switching strategy, we model the system as a repeated Bayesian game. We develop methods to assign attack actions to attacker types and generate realistic utilities based on expertise of security professionals.  For obtaining real-world attack data, we mine vulnerabilities in the National Vulnerability Database (NVD) and obtain utilities based on the Common Vulnerability Scoring System (CVSS).  We formulate an optimization problem which outputs a switching strategy that maximizes system security while accounting for switching costs.  The generated strategy is shown to be more effective than the state-of-the-art for a real-world application.  We also provide metrics that can be used to validate the robustness of switching strategies, absent in literature for multi-agent cyber-security systems.  Lastly, we propose the problem of identifying critical vulnerabilities and provide a solution.

The techniques in this paper are not limited to only web applications.  The attack actions mined from the security databases relate to all kinds of technologies, like operating systems, coding languages etc.  Hence, the modelling should be relevant to any software applications using the MTD architecture.  It would be interesting to see how effective they are in such scenarios.

Investigating the reward structure for a particular problem has helped design provably fast solvers in the physical security domains.  We believe this direction of research might help in developing faster solvers, alleviating the scalability problem of identifying critical vulnerabilities, for the cyber-security domain as well.




\section{Acknowledgments}
This work was partially supported by the grants from National Science Foundation (NSF-SFS-1129561) and the Center for Cybersecurity and Digital Forensics at Arizona State University.

\bibliographystyle{abbrv}
\bibliography{mtd}

\begin{thebibliography}{10}

\bibitem{amingradient}
K.~Amin, S.~Singh, and M.~Wellman.
\newblock Gradient methods for stackelberg security games.
\newblock {\em AAMAS}, 2016.

\bibitem{balzarotti2008saner}
D.~Balzarotti, M.~Cova, V.~Felmetsger, N.~Jovanovic, E.~Kirda, C.~Kruegel, and
  G.~Vigna.
\newblock Saner: Composing static and dynamic analysis to validate sanitization
  in web applications.
\newblock In {\em Security \& Privacy 2008. IEEE Symposium}, pages 387--401,
  2008.

\bibitem{gt-dyna}
K.~M. Carter, J.~F. Riordan, and H.~Okhravi.
\newblock A game theoretic approach to strategy determination for dynamic
  platform defenses.
\newblock In {\em ACM MTD Workshop, 2014}, MTD '14. ACM, 2014.

\bibitem{mtd-cn}
M.~Carvalho and R.~Ford.
\newblock Moving-target defenses for computer networks.
\newblock {\em Security Privacy, IEEE}, 12(2):73--76, Mar 2014.

\bibitem{bsg}
V.~Conitzer and T.~Sandholm.
\newblock Computing the optimal strategy to commit to.
\newblock In {\em Proceedings of the 7th ACM Conference on Electronic
  Commerce}, EC '06, pages 82--90, New York, NY, USA, 2006. ACM.

\bibitem{doupe2012enemy}
A.~Doup{\'e}, L.~Cavedon, C.~Kruegel, and G.~Vigna.
\newblock Enemy of the state: A state-aware black-box web vulnerability
  scanner.
\newblock In {\em USENIX Security Symposium}, 2012.

\bibitem{houmb2010quantifying}
S.~H. Houmb, V.~N. Franqueira, and E.~A. Engum.
\newblock Quantifying security risk level from cvss estimates of frequency and
  impact.
\newblock {\em JSS}, 83(9):1622--1634, 2010.

\bibitem{jones2015evaluating}
S.~Jones, A.~Outkin, J.~Gearhart, J.~Hobbs, J.~Siirola, C.~Phillips, S.~Verzi,
  D.~Tauritz, S.~Mulder, and A.~Naugle.
\newblock Evaluating moving target defense with pladd.
\newblock Technical report, Sandia National Labs-NM, Albuquerque, 2015.

\bibitem{surface}
P.~Manadhata.
\newblock Game theoretic approaches to attack surface shifting.
\newblock In {\em Moving Target Defense II}, volume 100 of {\em AIS}, pages
  1--13. Springer New York, 2013.

\bibitem{mccumber1991information}
J.~McCumber.
\newblock Information systems security: A comprehensive model.
\newblock In {\em Proceedings of the 14th National Computer Security
  Conference}, 1991.

\bibitem{mell2007cvss}
P.~Mell, K.~Scarfone, and S.~Romanosky.
\newblock Cvss v2 complete documentation, 2007.

\bibitem{okhravi2014finding}
H.~Okhravi, T.~Hobson, D.~Bigelow, and W.~Streilein.
\newblock Finding focus in the blur of moving-target techniques.
\newblock {\em Security \& Privacy, IEEE}, 12(2):16--26, 2014.

\bibitem{dobss}
P.~Paruchuri, J.~P. Pearce, J.~Marecki, M.~Tambe, F.~Ordonez, and S.~Kraus.
\newblock Playing games for security: An efficient exact algorithm for solving
  bayesian stackelberg games.
\newblock In {\em AAMAS, 2008}, pages 895--902, 2008.

\bibitem{armor}
J.~Pita, M.~Jain, J.~Marecki, F.~Ord{\'{o}}{\~{n}}ez, C.~Portway, M.~Tambe,
  C.~Western, P.~Paruchuri, and S.~Kraus.
\newblock Deployed {ARMOR} protection: the application of a game theoretic
  model for security at the los angeles international airport.
\newblock In {\em AAMAS 2008, Industry and Applications Track Proceedings},
  pages 125--132, 2008.

\bibitem{Prakash15EGA}
A.~Prakash and M.~P. Wellman.
\newblock Empirical game-theoretic analysis for moving target defense.
\newblock In {\em ACM MTD Workshop, 2015}, 2015.

\bibitem{nyt}
J.~Silver-Greenberg, M.~Goldstein, and N.~Perlroth.
\newblock {JPMorgan Chase Hacking Affects 76 Million Households}.
\newblock In {\em The New York Times}, 2014.

\bibitem{Sinha16a}
A.~Sinha, T.~Nguyen, D.~Kar, M.~Brown, M.~Tambe, and A.~X. Jiang.
\newblock From physical security to cyber security.
\newblock {\em Journal of Cybersecurity}, 2016.

\bibitem{taguinodtoward2015}
M.~Taguinod, A.~Doup\'e, Z.~Zhao, and G.-J. Ahn.
\newblock {Toward a Moving Target Defense for Web Applications}.
\newblock In {\em Proceedings of 16th IEEE IC-IRI}, 2015.

\bibitem{van2013flipit}
M.~Van~Dijk, A.~Juels, A.~Oprea, and R.~L. Rivest.
\newblock Flipit: The game of ``stealthy takeover''.
\newblock {\em Journal of Cryptology}, 26(4):655--713, 2013.

\bibitem{von2010market}
H.~Von~Stackelberg.
\newblock {\em Market structure and equilibrium}.
\newblock Springer SBM, 2010.

\bibitem{wicaksono2008piecewise}
D.~S. Wicaksono and I.~Karimi.
\newblock Piecewise milp under-and overestimators for global optimization of
  bilinear programs.
\newblock {\em AIChE Journal}, 54(4):991--1008, 2008.

\bibitem{wichers2013owasp}
D.~Wichers.
\newblock Owasp top-10.
\newblock {\em OWASP}, 2013.

\bibitem{winterrose2014adaptive}
M.~Winterrose, K.~Carter, N.~Wagner, and W.~Streilein.
\newblock Adaptive attacker strategy development against moving target cyber
  defenses.
\newblock {\em arXiv:1407.8540}, 2014.

\bibitem{zhuang2014towards}
R.~Zhuang, S.~A. DeLoach, and X.~Ou.
\newblock Towards a theory of moving target defense.
\newblock In {\em ACM MTD Workshop, 2014}, pages 31--40. ACM, 2014.

\end{thebibliography}
\end{document}